\documentclass[reqno,12pt]{article}
\usepackage{amsmath,amsfonts,amssymb,amsthm,amstext,amscd,eucal}
\usepackage[all]{xy}

\makeatletter \@addtoreset{equation}{section}

\makeatletter\renewcommand\section{\@startsection {section}{1}{\z@}%
                                   {-3.5ex \@plus -1ex \@minus -.2ex}
                                   {2.3ex \@plus.2ex}%
                                   {\normalfont\large\bfseries}}
\renewcommand\subsection{\@startsection{subsection}{2}{\z@}%
                                     {-3.25ex\@plus -1ex \@minus -.2ex}%
                                     {1.5ex \@plus .2ex}%
                                     {\normalfont\bfseries}}

\newcommand{\be}{\begin{equation}}
\newcommand{\ee}{\end{equation}}
\newcommand{\beq}{\begin{eqnarray}}
\newcommand{\eeq}{\end{eqnarray}}
\newcommand{\bea}{\begin{eqnarray}}
\newcommand{\eea}{\end{eqnarray}}

\def\[{\left [}
\def\]{\right ]}
\def\({\left (}
\def\){\right )}

\def\CN{{\cal N}}

\def\r2{\sqrt{2}}

\newcommand{\U}{\mathrm{U}}

\def\sst#1{{\scriptscriptstyle #1}}

\def\1{{\sst{(1)}}}

\def\CM{{\cal M}}
\def\CN{{\cal N}}
\def\CO{{\cal O}}

\newcommand{\bbibitem}[1]{\bibitem{#1}\marginpar{#1}}

\def\Label#1{\label{#1}%
  \smash{\hbox to0pt{\raise1ex\hbox{\tiny[#1]}\hss}}}
\def\noLabels{\let\Label=\label}
\def\nobbibitem{\let\bbibitem=\bibitem}

\begin{document}

\noLabels 
\nobbibitem 

\begin{titlepage}



\begin{center}

{\Large{\bf Small Black holes vs horizonless solutions in AdS}} \vfil \vspace{3mm}

{\bf Joan Sim\'on\footnote{e-mail: J.Simon@ed.ac.uk}}
\\

\vspace{8mm}

\bigskip\medskip
\begin{center} 
{\it School of Mathematics and Maxwell Institute for Mathematical Sciences,\\
King's Buildings, Edinburgh EH9 3JZ, United Kingdom}\\
\end{center}
\vfil

\end{center}
\setcounter{footnote}{0}

\begin{abstract}
\noindent It is argued that the appropriate macroscopic description of half-BPS mesonic chiral operators in generic $d=4$ ${\cal N}=1$ toric gauge theories is in terms of the geometric quantization of smooth horizonless configurations. The relevance of different ensemble macroscopic descriptions is emphasized : lorentzian vs euclidean configurations as (semiclassical) microstates vs saddle points in an euclidean path integral.
\end{abstract}
\vspace{0.5in}

\end{titlepage}
\renewcommand{\baselinestretch}{1.05}  


\section{Introduction}

Given a microscopic degeneracy $d_{\text{micro}}(Q_i)$ of states carrying charges $Q_i$, one would like to have a macroscopic derivation of the same result
\begin{equation*}
  d_{\text{micro}}(Q_i) = d_{\text{macro}}(Q_i)\,,
\end{equation*}
in terms of manifest gravitational degrees of freedom to improve our understanding on the nature of quantum gravity and the emergence of spacetime.

In string theory, it is generically expected that higher order corrections in both $\alpha^\prime$ and $g_s$ must play an important role in establishing an accurate connection. Even without their inclusion, progress was achieved with the construction of classical configurations carrying the same charges as black holes, but without a horizon. The latter helped to develop the fuzzball proposal (see \cite{first} for original work and \cite{fuzzball} for reviews).


This program has been made explicit for supersymmetric small black holes. In \cite{Sen:2009bm} , it was argued that in a given duality frame, either the system allows $\alpha^\prime$ corrections to generate a horizon, or the system allows the existence of smooth horizonless configurations. Either way, the macroscopic description reproduces the microscopic entropy, either through euclidean path integral considerations or through geometric quantization \cite{Crnkovic:1986ex}.

In this note, we extend the arguments given in \cite{Sen:2009bm} to small black holes in $AdS_5$.
More precisely, we consider the macroscopic description of states belonging to the chiral ring in $\CN=4$ SYM and the half-BPS mesonic chiral sector of generic $\CN=1$ toric quiver theories with conformal dimension $\Delta \sim N^2$. The microscopic partition functions indicate that the growth in the entropy in these sectors  is not fast enough to generate a macroscopic classical horizon. This is consistent with the existence of singular configurations with horizon at the singularity \cite{sabra,buchel,kim}. It will be argued that $\alpha^\prime$ corrections can not generate a horizon, because the charge dependence of the entropy in these sectors is not compatible with the existing scaling symmetry of the gravitational macroscopic classical lagrangian description. Exactness in the partition function for non-vanishing $g_s$ suggests this conclusion is not modified by the full quantum macroscopic action. 

Since singular configurations do not belong to the configuration space of the macroscopic theory, these states must allow a description in terms of the geometric quantization of a set of smooth horizonless configurations, naturally generalizing  the picture emerging in the 1/2 BPS sector of $\CN=4$ SYM \cite{Lin:2004nb}, where matching with the gauge theory was achieved in \cite{Grant:2005qc,Maoz:2005nk}.

A brief discussion on the implications of these statements for large black holes is included at the end. The different role that different ensembles play is emphasized : on the one hand, the existence of scaling solutions \cite{bena-scaling,scaling} in the lorentzian (microcanonical) formulation, including the description of the interior of the black hole, and the relevance of both $\alpha^\prime$ and $g_s$ corrections to reproduce the microscopic entropy through geometric quantization \cite{deBoer:2008zn,deBoer:2009un}; on the other hand, the existence of smooth euclidean black hole configurations, for which such interior is removed, being saddle points of an euclidean path integral with suitable boundary conditions describing a canonical ensemble. This distinction is consistent with the lorentzian vs euclidean formulations of the AdS/CFT correspondence.


%




\section{The scaling argument in AdS}

In this section, it is first reviewed the existence of an scaling symmetry for the classical type IIB string theory action \cite{senscaling,Sen:2009bm}. The latter will be used to argue that $\alpha^\prime$ corrections can not generate horizons for the subset of asymptotically $AdS_5$ configurations considered in this note.

The scaling symmetry consists of shifting the dilaton $\phi$ by a constant $\log\lambda^{-1}$, keeping all other Neveu-Schwarz-Neveu-Schwarz (NS-NS) sector fields invariant and multiplying the Ramond-Ramond (RR) sector fields by $\lambda$. The overall transformation scales the action by $\lambda^2$.

The argument below requires to know the scaling properties of the different charges. Since magnetic charges are directly related to the magnetic components of the different fields, we infere $Q_{NSNS}^{\text{mag}}$ remains invariant, while $Q_{RR}^{\text{mag}}$ is multiplied by $\lambda$. On the other hand, since electric charges are related to the derivative of the action with respect to the electric field, $Q_{NSNS}^{\text{el}}$ scales like $\lambda^2$, while $Q_{RR}^{\text{el}}$ does so linearly.

To explore the generation of horizons by $\alpha^\prime$ corrections, we will study the entropy. Since the latter can be computed using Wald's formula \cite{wald} for any given classical configuration, we conclude it scales quadratically under the scaling symmetry :
\begin{equation}
  S_{\textrm{grav}} \left(\lambda\,Q_{RR},\,\lambda^2\,Q_{NSNS}^{\text{el}},\,Q_{NSNS}^{\text{mag}}\right) = \lambda^2\,
  S_{\textrm{grav}}\left(Q_{RR},\,Q_{NSNS}^{\text{el}},\,Q_{NSNS}^{\text{mag}}\right)\,.
 \label{eq:sentropy}
\end{equation}

As pointed out in \cite{Sen:2009bm}, the above relation assumes the entropy for a given set of charges to be independent of the asymptotic value of the dilaton and the moduli arising in the RR sector. For asymptotically $AdS_5$ configurations with constant dilaton and non-trivial RR flux at infinity, consistency of the scaling transformation of the different dynamical fields requires to keep
the radius of $AdS_5$
\begin{equation*}
  L^4 =4\pi\,g_s N\,l_s^4 \quad \text{invariant} \quad \Rightarrow \quad g_s\to \lambda^{-1}\,g_s\,,\,\,\,\, N\to \lambda\,N
\end{equation*}
i.e. to keep the 't Hooft coupling invariant. This is the scaling symmetry that will be used in this note.

\subsection{Half-BPS sector in $\CN=4$ SYM}

The $\CN=4$ SYM partition function for half-BPS states 
\begin{equation}
  Z(\nu,\,q) = \prod_{n=0}^\infty \frac{1}{1-\nu\,q^n}
\end{equation}
is not affected by quantum corrections, and as such it can be computed in the free theory limit and extrapolated to strong coupling. Here, $\nu$ is the chemical potential for the number of D-branes $N$, $q=e^{-\beta}$ with $\beta$ the chemical potential dual to R-charge $n=J$.

This sector of the theory has no phase transition at large $N$ \cite{Kinney:2005ej}. The entropy at very small chemical potential $\beta$ with conformal dimension $\Delta = J\sim N^2$ is \cite{david,Kinney:2005ej} \footnote{It is well known that states with $\Delta \sim \CO(1)$ correspond to perturbative gravitons, those with $\Delta\sim \CO (N)$ to (dual) giant gravitons and those with $\Delta\sim \CO (N^2)$ to small black holes with suitable gravitational description.}
\begin{equation}
  S_{\textrm{1/2-BPS}} \propto N \log N\,.
\end{equation}
This result captures  the large temperature behavior of N harmonic oscillators plus an $1/N!$ statistical factor.


Given the exact nature of the partition function, we can estimate the size of an stretched horizon by comparing the field theory entropy with the Bekenstein entropy :
\begin{equation}
  S_\text{grav} =  S_{\text{1/2-BPS}} \sim N^2\,\left(\frac{\rho_h}{L}\right)^3 \quad \Rightarrow \quad \frac{\rho_h}{L}\ll 1\,.
 \label{eq:shorizon}
\end{equation}
Thus, for half-BPS states of conformal dimension $\Delta\sim N^2$, for which we expect to have a reliable gravitational description in terms of asymptotically $AdS_5\times S^5$ spacetimes, this estimation confirms their degeneracy is not large enough to generate a macroscopic horizon, not even a string scale size one.

\paragraph{Gravity description \& higher order corrections :} There exist two classical descriptions in gravity with these quantum numbers and symmetries :
\begin{itemize}
\item The extremal BPS limit of non-extremal single R-charged black holes \cite{sabra}. This is a singular configuration, the so called superstar \cite{myers}, in which the zero size horizon coincides with a null naked singularity.
\item There exist classical horizonless configurations with the same conserved charges as the superstar, which can be smooth if appropriate boundary conditions are chosen \cite{Lin:2004nb}.
\end{itemize}

It is important to establish whether higher order $\alpha^\prime$ corrections to supergravity can generate a horizon, since our previous estimation only included the lowest order contribution to Wald's formula. To answer this question, notice that the gauge theory entropy dependence on the angular momentum $J$, which is an electric NS-NS charge, is proportional to $\sqrt{J}\log \sqrt{J}$. Thus, under the scaling symmetry \eqref{eq:sentropy}, the microscopic entropy satisfies,
\begin{equation}
  S_{\text{1/2-BPS}}(\lambda^2\,J) \neq \lambda^2\,S_{\textrm{grav}}(J)\,.
\end{equation}
This scaling is not compatible with the one derived from a macroscopic (gravitational) description of the system; thus, $\alpha^\prime$ corrections to type IIB will not be able to generate any horizon and the superstar configuration will remain singular.

Following the philosophy described in \cite{Sen:2009bm}, one reaches the conclusion that the superstar is {\it not} a proper {\it classical} solution to the equations of motion. Thus, it does not belong to the physical configuration space. On the other hand, we know of the existence of a classical moduli space of smooth horizonless configurations in  \cite{Lin:2004nb}, whose geometric quantization \cite{Crnkovic:1986ex} reproduces the Hilbert space describing the 1/2 BPS sector in the gauge theory \cite{Maoz:2005nk}. One is thus left to conclude that the appropriate macroscopic description for these half-BPS states is in terms of the latter. 

\paragraph{Dependence on the ensemble : } though the previous discussion is natural from a microscopic point of view, it was realized in \cite{babel} that the structure of typical states in such ensemble did not reproduce the superstar geometry. The reason is that many of the operators counted above do not describe a bound state of a given number $N_c$ of giant gravitons. This can be relevant because the singular superstar is characterized by a non-trivial flux $q/L^2=N_c/N$, accounting for such number. When such constraint is taken into account\footnote{The procedure is not unique, but for our purposes here we will not enter into the details of this subtlety \cite{mueck}.}, one is able to match the superstar geometry as the one corresponding to a typical state in the constrained ensemble. The entropy was computed in \cite{babel} to be
\begin{equation}
  S_{\text{sstar}} = - N\log \frac{\omega^\omega}{(1+\omega)^{(1+\omega)}}\,, \quad \quad \omega= \frac{N_c}{N}\,.
  \label{sstarent}
\end{equation}
Notice that $\omega$ remains invariant under the scaling symmetry described above since $N_c$ scales like $\lambda\,N_c$, due to the fact that it is linear in the RR 5-form flux. 

The R-charge of the superstar satisfies $J=\omega\,N^2/2$. Thus, the entropy computed in the gauge theory satisfies $S_{\text{sstar}}=\sqrt{J}\,f(\omega)$ and as such, it still scales linearly in $\lambda$. Thus, even working in this set of ensembles, the microscopic entropy will never scale as the macroscopic entropy computed from the type IIB macroscopic action. Our conclusion remains unchanged.

\subsection{1/4 and 1/8 BPS sectors in $\CN=4$ SYM}

The $\CN=4$ SYM partition functions in the 1/4 and 1/8 BPS sectors suffer from a discontinuity at the free theory point, but once the coupling is turned on, they are claimed to be exact \cite{Kinney:2005ej}. Thus, one can compute them at weak coupling and once again extrapolate to strong coupling. Another new feature in these sectors is the existence of a second order phase transition at large N.

All required to extend our previous arguments to these sectors is to compute the entropy scaling with $N$, check whether there is no macroscopic stretched horizon consistent with that scaling and to argue the microscopic entropy scaling is not consistent with the scaling symmetry required by the macroscopic lagrangian, ensuring $\alpha^\prime$ corrections will not generate such horizon either.

As explained in \cite{Kinney:2005ej}, operators with conformal dimension $\Delta\sim N^2$ and having equal $\U(1)$ R-charges, in the large N limit,  belong to a phase in which the entropy scales as
\begin{equation}
  S_{\textrm{gauge}}(J,\,N)\sim N\log N \quad \text{with} \quad J\sim N^2
\end{equation}
The existence of a macroscopic, or even string scale, stretched horizon requires $S_{\text{gauge}} \sim N^2$, as seen from \eqref{eq:shorizon}. Thus, the entropy is not large enough to generate a horizon in classical gravity in both BPS sectors. 

As in the half-BPS case, the scaling of the gauge theory entropy under the existing scaling symmetry in the macroscopic description satisfies
\begin{equation}
  S_{\textrm{gauge}}(\lambda^2J,\,\lambda N) \neq \lambda^2\,S_{\textrm{grav}}(J,\,N)\,.
\end{equation}
Thus, the identification $S_{\textrm{gauge}}(J,\,N)=S_{\textrm{grav}}(J,\,N)$ is incompatible with such symmetry. As before, we conclude that such $\alpha^\prime$ corrections will not be able to generate a horizon. This conclusion will also hold for $g_s$ corrections due to the exact character of the gauge theory partition function once $g_s\neq 0$\footnote{The fact that these states do not correspond to black holes was already pointed out in \cite{david}.}. 

\paragraph{Gravity evidence :} there exist analogous singular configurations to the 1/2 BPS superstar in the 1/4 and 1/8 BPS sectors, the extremal BPS limits of the corresponding multi R-charged black holes in $AdS_5$ \cite{sabra}. There has been some work searching for regular horizonless configurations asymptoting to $AdS_5\times S^5$ preserving the right amount of supersymmetry and having the appropriate global symmetries \cite{aristos1,donos2,milanesi,microstates,olegbubb}. Even the characterization of the classical moduli space for such configurations is difficult due to the non-linearity of the differential equations describing the system (at least in the coordinates that have been used so far) and the global regularity requirements that must be imposed on them. 

In the 1/4 BPS case, such configurations are locally characterized by a four dimensional K\"ahler manifold
\begin{equation}
  ds^2 = -h^{-2}\,(dt+\omega)^2 + h^2\left(\frac{2}{Z+1/2}\partial_a\overline\partial_b K dz^ad\overline{z}^b + dy^2\right) + y\left(e^G\,d\Omega_3^2 + e^{-G}\,d\psi^2\right)\,,
\end{equation}
with K\"ahler potential $K(y,z^a,\overline{z}^b)$ satisfying a non-linear Monge-Ampere equation \cite{microstates} :
\begin{equation}
 \textrm{det} \partial_a\overline\partial_b K = \frac{y}{4}\left(Z+\frac{1}{2}\right)\,e^{-y\partial_y K}\,, \quad\quad Z = -\frac{1}{2}y\partial_y\left(y^{-1}\partial_y K\right)\,.
\label{eq:ampere}
\end{equation}
The entire configuration is determined in terms of $K$. Regularity requires a whole set of boundary conditions :  first, from the behavior of the K\"ahler potential in the deep interior of the solution $(y\to 0)$ \cite{microstates}, giving rise to some droplet picture, and second, from some holomorphicity condition constraining the shape of the latter \cite{olegbubb}, derived by matching probe calculations with the supergravity analysis. It is these last conditions that render the solution to the Monge-Ampere equation \eqref{eq:ampere} unique,  in perturbation theory \cite{olegbubb}. 

In the 1/8 BPS case, supersymmetry requires the metric to locally satisfy \cite{microstates} 
\begin{equation}
  ds^2 = -e^{2\alpha}\,(dt+\omega)^2 + e^{-2\alpha}\,h_{ij}dx^idx^j + e^{2\alpha}\,d\Omega_3^2\,,
 \label{eq:18bps}
\end{equation}
where $e^{4\alpha} = -3R^{-1}$, $R$ being the scalar curvature of the six dimensional manifold that also satisfies the non-linear equation \cite{kim0}
\begin{equation}
  \Box_6 R = -R_{ij}\,R^{ij} + \frac{1}{2}\,R^2
\end{equation}
Despite some analysis of regularity conditions given in \cite{microstates}, we are not aware of a complete attempt to investigate the global regularity of these configurations along the lines of \cite{olegbubb}, attempting to match any such configuration with some classical probe profile.

The arguments given above strongly suggest the existence of such configurations, a subset of them being smooth and whose geometric quantization should give rise to a Hilbert space matching the gauge theory description. This would provide with a macroscopic derivation of the results reported in \cite{sparks-Z} obtained through geometric quantization of the classical moduli space of dual giant gravitons in the probe approximation. The latter are obtained at weak, but non-vanishing string coupling and match the microscopic gauge theory partition function. Since these are independent of $g_s$, once the latter is different from zero, we expect an extension of the results reported in \cite{Maoz:2005nk} for LLM configurations to these 1/4 and 1/8 configurations.

\subsection{Mesonic half-BPS sector in $\CN=1$ toric gauge theories}

The main conclusions of the previous sections can be extended to the mesonic half-BPS sector in
generic $\CN=1$ toric quiver gauge theories dual to $AdS_5\times\CM^5$. 

Either by quantizing the classical moduli space of dual giants in these backgrounds \cite{beasley,minwalla,sparks-Z}  or by directly solving the combinatorial problem of counting the number of mesonic gauge invariant supersymmetric operators carrying some R-charge $J$ \cite{Feng:2007ur}, the partition functions in these sectors can be computed :
\begin{equation}
  Z(\nu,\,t) = \prod_{J=0}^\infty\frac{1}{\left(1-\nu\,q^J\right)^{a_J}}\,,
\end{equation}
where $a_J$ stands for the degeneracy of single trace operators carrying charge $J$. 

Even though these partition functions do not include any fermionic contribution, they count the number of giant gravitons (or dual giant gravitons), by construction. These are the sources for the naked singularity existing in the supergravity configurations carrying the same amount of mass  \cite{buchel,kim}, extending the analysis and interpretation originally done in \cite{myers} in the context of $\CN=4$ SYM and $AdS_5\times S^5$. Thus, this is the appropriate partition function to compare with the existing gravity configurations.

It was shown in  \cite{Feng:2007ur} and \cite{Balasubramanian:2007hu}, using algebraic and geometric methods, respectively, that the $J$ functional dependence of the dominant entropy contribution is, in the large charge limit, {\it universal}, in the sense that its scaling does not depend on the manifold $\CM_5$ but rather only on the complex dimension of the cone manifold over $\CM_5$.

The large charge limit forces $q\to 1$ (small chemical potential), but the behaviour of $\nu$ determines distinct phases with different scaling entropy \cite{Feng:2007ur} :
\begin{itemize}
\item[1.] When $q,\,\nu\to 1$, the entropy is
\begin{equation}
  S_{\textrm{gauge}}(J) \sim \left(V_3 \zeta(4)\right)^{1/4}\, J^{3/4}
\end{equation}
where $V_3$ is the normalised Einstein-Sasaki volume.
\item[2.] When $q\to 1$ and $\nu\to 0$, the entropy is proportional to $N\log N$
\begin{equation}
  S_{\textrm{gauge}}(J)\sim N\log N
\end{equation}
\end{itemize}
The second phase requires $N\ll J^{3/4}$, and the transition occurs at $N\sim J^{3/4}$.

Since we are interested in operators of conformal dimension $\Delta\sim N^2$, the system will be in the second phase. Given this microscopic gauge theory entropy, the estimation for the stretched horizon $\rho_h$ in \eqref{eq:shorizon} still applies here \cite{Balasubramanian:2007hu} :
\begin{equation}
S_{\textrm{grav.}} \sim \frac{\textrm{Vol}(X)}{l_p^5}\,
\frac{\rho_h^3}{l_p^3} \quad \Rightarrow \quad \left(\frac{\rho_h}{L}\right)^3  \sim
\frac{a_{\mathcal{N}=1}}{a_{\mathcal{N}=4}}  \,
\frac{S_{\textrm{gauge.}}}{N^2}\,,
\end{equation}
where we used \cite{sparks-volume} $a_{\mathcal{N}=1}\,\textrm{Vol}(X) = a_{\mathcal{N}=4}\,\textrm{Vol}(S^5)$ and identified $S_{\textrm{grav.}}=S_{\textrm{gauge}}$. Thus, if the gauge theory entropy does not grow like $N^2$, in the large N limit, the stretched horizon is not macroscopic, and not even string scale, since $g_s\,N$ is large but finite.

One may again worry about the role of quantum corrections in the macroscopic side of our description, but the scaling of the microscopic entropy $S_{\textrm{gauge}}$ is again not compatible with the macroscopic scaling \eqref{eq:sentropy} required to be derivable from a type IIB macroscopic lagrangian. Thus, higher order $\alpha^\prime$ corrections to the singular supergravity configurations \cite{buchel,kim} will remain singular. This discussion suggests that for {\it all} these theories and their gravity duals, there should exist a classical moduli space of horizonless smooth configurations whose quantization should reproduce the entropy counting from the gauge theory. Actually, following \cite{david1}, the above claim can be extended to any cone over an Einstein-Sasaki manifold $\CM^5$, and not just to toric cones.

The gravity evidence in favour of our interpretation is very similar to the one explained in the 1/8 BPS sector in $\CN=4$ SYM. Both cases have the same amount of supesymmetry, thus locally the metric will allow a similar decomposition to the one in \eqref{eq:18bps}. The asymptotic boundary conditions will be different though : the asymptotic background should already break $1/4$ of the supersymmetry, and it is the extra R-charge that breaks 1/2 of the remaining supersymmetry. It is because of this fact that we expect a reduction in the difficulty of the non-linear equations once the appropriate supersymmetry is imposed. Some preliminary work was done in \cite{milanesi2} for half-BPS states in $AdS_5\times Y^{(p,q)}$, but no regularity analysis was performed.

Once again, we are confident these configurations exist, and their geometric quantization will reproduce the mesonic partition function. This would be consistent with the geometric quantization of the classical moduli space of dual giant gravitons in the probe approximation \cite{sparks-Z}. Given the isomorphism between this Hilbert space of dual giants and of giant gravitons, and the non-renormalization in $g_s$, we do expect supergravity to capture this same information.

\section{Discussion on large black holes}

It is important to extend the lessons above from small to large black holes. Different formalisms may suggest different answers, but it will be argued this is not necessarily the case. Ensemble wise, one such formalism is intrinsically canonical in nature and is based on euclidean path integral considerations; the second one is microcanonical and based on lorentzian geometries.

For small black holes that do not get corrected in $\alpha^\prime$, both the lorentzian and euclidean versions of the hole are singular configurations. Thus, one is instructed \cite{Sen:2009bm} to consider the description in terms of horizonless configurations (if they exist). These being lorentzian geometries, the question remains as for a possible euclidean reformulation of this claim\footnote{There exists the possibility of having singular configurations becoming smooth only after the inclusion of $g_s$ corrections in a given duality frame.}. If the black hole gets corrected in $\alpha^\prime$, euclidean path integral considerations can account for the entropy. This does {\it not} exclude though the existence of horizonless configurations when $g_s$ corrections are also included in the lorentzian set-up.

For large black holes, the last point is already relevant at lowest order in the classical action : lorentzian black hole solutions exist, with curvature singularities in their deep interior. As such, they are generically {\it not} solutions to the classical equations of motion. Their euclidean continuations, however,  become saddle points in an euclidean path integral approach and are smooth (when suitable boundary conditions are imposed), at the expense of removing the interior of the geometry. It is this distinction that is at the core of the fuzzball proposal.

By construction, such saddle points provide the dominant contributions to the 
macroscopic entropy $d_{\text{macro}}$. Thus, the euclidean formalism knows about the total number of states, but information about the individual microstates seems to be lost (or it is not manifest in our current understanding). Notice this remark is consistent with the different available formulations of the AdS/CFT correspondence. There is an euclidean path integral formulation that allows to compute the partition function, and extract the total number of states, but we also know that the use of lorentzian geometry is essential to capture the difference between microstates through the expectation values of the different gauge invariant operators encoded in the boundary fall-off conditions of the different bulk fields \cite{lor-ads/cft}. As it is well-known \cite{farey-tale}, such saddle points are {\it not} a semiclassical description of states with well-defined mass and spin in a quantum gravity Hilbert space.

To emphasize the dicotomy of these descriptions, let us review Sen's proposal to account for the macroscopic degeneracy of supersymmetric black holes having an $AdS_2$ throat \cite{Sen:2008vm,Sen:2009vz}
\begin{equation}
  d_{\text{macro}}(Q) = \sum_s \sum_{
      Q_i,\, Q_{\text{hair}}\atop \sum_{i=1}^s Q_i + Q_{\text{hair}} = Q} \Big\{\prod_{i=1}^s d_{\text{hor}}(Q_i)\Big\}\,d_{\text{hair}}(Q_{\text{hair}};\,\{Q_i\})\,.
 \label{eq:senp}
\end{equation}
The s-th term represents the contribution from an s-centered black hole configuration; $d_{\text{hor}}(Q_i)$ stands for the degeneracy associated with the horizon of the i-th black hole center carrying charge $Q_i$; and $d_{\text{hair}}(Q_{\text{hair}};\,\{Q_i\})$ stands for the hair degeneracy, i.e. smooth black hole deformations supported outside the horizon and sharing the same asymptotics. 

Sen's prescription uses a mixture of formulations. Indeed, whereas the contribution from the degrees of freedom localised at the horizon is captured by an euclidean path integral, both the contribution from horizonless configurations, through geometric quantization, and hair modes employ entirely lorentzian methods. Thus, for large black holes, there may appear to be a tension between both ensemble descriptions
\begin{itemize}
\item[1.] From an euclidean path integral perspective with standard boundary conditions, the contribution from the euclidean continuations of the horizonless configurations (if any) must be subleading.
\item[2.] From a lorentzian geometry point of view, recent attempts to account for the microscopic entropy through geometric quantization of scaling solutions \cite{deBoer:2008zn,deBoer:2009un}, in the supergravity approximation, showed their contribution was subleading.
\end{itemize}

Even though this may suggest that the contribution from the quantization of the classical moduli space of horizonless configurations is always subleading for large black holes, both remarks are neither conclusive nor exclusive :
\begin{itemize}
\item[1.] If euclidean continuations of horizonless configurations do contribute at all to the standard euclidean path integral formulation, their role may be quite different, since their "thermal" circles are non-contractible, whereas any black hole contribution from $s\geq 1$ in \eqref{eq:senp} will involve contractible circles. Anyway, the precise rules to deal with these configurations in an euclidean path integral are currently not clear, and they may require to include complex metrics preserving the reality of the action.
\item[2.] Higher order corrections, both in $\alpha^\prime$ and $g_s$, should increase the degeneracy computed in \cite{deBoer:2009un} in the lorentzian geometry set-up.
\end{itemize}

Interestingly, for half-BPS small black holes in $AdS_5$, it was already shown in \cite{babel}, that the singular small black hole (the superstar) did correspond to the typical microstate of the system. The matching involved a coarse-graining in the phase space of the quantum mechanics describing the dual system of free fermions (see \cite{cg-oleg} for a similar discussion in the D1-D5 system). After such coarse-graining has ocurred, the information regarding the individual microstates is partially lost; entropy is thus generated and the spacetime geometry becomes singular \cite{donus}. This procedure does attempt to establish a bridge between both formulations. Later, it was emphasized how difficult it is to tell apart such thermal density matrix from a given typical microstate \cite{masakivijay,mukundus}, justifying why lorentzian black holes are such good approximations for semiclassical physics considerations. But, despite capturing most of the (classical) information, these states are obviously different as quantum states: what remains true though, is the capture of the degeneracy of microstates by the thermal description.

The idea is that a similar picture can emerge for large black holes. There is no reason to expect that such scenario can be checked at lowest order in supergravity. And indeed, the preliminary results reported in \cite{deBoer:2009un} do confirm this expectation. It does teach us that scaling solutions may well play a similar role to the horizonless configurations advocated in this work, but that they require
higher order corrections, both in $\alpha^\prime$ and $g_s$ to achieve this goal.

It is clear that clarifying these issues is of primarily importance to unify recent developments in the fuzzball proposal context with more traditional approaches to black hole entropy based on euclidean path integral formulations. It is also apparent that similar considerations will arise in any microscopic understanding of cosmological horizons and Rindler space.



\section*{Acknowledgements}
J.S. would like to thank David Berenstein, Jan de Boer, Masaki Shigemori, Ashoke Sen and Erik Verlinde for useful discussions, and especially Ashoke Sen for encouragement. This work was partially supported by the Engineering and Physical Sciences Research Council [grant number EP/G007985/1] and by the Science and Technology Facilities Council [grant number Council ST/G000425/1].


\providecommand{\href}[2]{#2}\begingroup\raggedright

\endgroup

\end{document}